\documentclass[aps,pra,floatfix,twocolumn,superscriptaddress]{revtex4-1}
\usepackage[final]{graphicx}
\usepackage{times,amsmath,amssymb}
\usepackage{epsfig,color}
\usepackage{hyperref}
\usepackage[caption = false]{subfig}
\usepackage{thumbpdf,enumerate}
\usepackage{booktabs}
\usepackage{bbm} 

\newcommand{\Tr}{{\rm Tr}}

\def\Tr{\hbox{Tr}} 
\usepackage{pifont}

\begin{document}

\title{Reversibility in one-dimensional quantum cellular automata in the presence of noise}

\author{Federico Centrone}
\affiliation{Dipartimento di Fisica, Sapienza Universit\`a di Roma , Piazzale Aldo Moro 5, 00185 Rome, Italy.}

\author{Camillo Tassi}
\affiliation{Dipartimento di Matematica e Fisica, Universit\`{a} degli Studi Roma Tre, Via della Vasca Navale 84, 00146, Rome, Italy}

\author{Alessio Serafini}
\affiliation{Department of Physics and Astronomy, University College London, Gower Street, London WC1E 6BT, UK}

\author{Marco Barbieri}
\affiliation{Dipartimento di Scienze, Universit\`{a} degli Studi Roma Tre, Via della Vasca Navale 84, 00146, Rome, Italy}

\begin{abstract}
We consider a class of noisy, one-dimensional quantum cellular automata that allow one to shift from unitary dynamics to completely positive maps, 
and investigate the notion of reversibility in such a setting. To this aim, we associate an approximate reverse automaton to each noisy 
automaton, and assess its effect, and we define an irreversibility time based on the distance from the maximally mixed state, which is shown to be the only attractor of the automaton map in the presence of dephasing.
\end{abstract}

\maketitle


Time is what is measured by a clock, which is anything performing a periodic motion. The periodicity is fundamental to quantify the passage of time, but time itself is revealed already as soon as anything moves at all. However, a clock does not give any preferred direction to the dynamic so, in principle, the notions of past and future might be regarded as those of \textit{left} and \textit{right}. Surprisingly, they are when dealing with isolated microscopic systems. 

On the other hand, in our daily observation of Nature,this symmetry  is quite exceptional. Imagine, in fact, to observe the Universe as time flows backwards. You would immediately notice that something is wrong and unnatural.  Most of natural processes occurs \textit{spontaneously}, giving a precise direction to the stream of time ~\cite{bib:timearrow1, bib:timearrow2,eddington}. The first attempt to face the problem of the emergence of an \textit{arrow of time} in macroscopic system was held by Ludwig Boltzmann in his unification of mechanics and thermodynamics \cite{bib:Boltzmann}. In his picture, a macroscopic system spend most of its time in his most probable configuration, which has the highest entropy value. The spontaneous evolution of the system towards this configuration is just due to its initial conditions. If the initial state is at the equilibrium, on average the system has no dynamics, hence time stops.

For a many body system, if the various parts composing it starts the dynamics in an initial uncorrelated state, they will begin to develop correlations through their interaction, producing an average increase in entropy and setting a precise direction to the arrow of time. This has been confirmed in various experiments in both classical \cite{andrieux} and quantum mechanics \cite{paternostro,hofman}, establishing a way to manipulate and control the flow of time operating on the system's initial conditions. 

Quantum information theory had a remarkable impact on our understanding of irreversible phenomena. Besides the increasing interest in the development of quantum devices based on reversible \textit{conservative logic} \cite{peres,deVos}, the problem of reversibility is of striking importance in the foundations of the theory itself. While the evolution of a closed quantum system, expressed by a unitary transformation, is deterministic and invertible, the coupling with an unknown environment injects an uncontrolled element into the dynamics. The interaction with a thermal bath, in fact, establishes correlations with a system whose state is unknown, which inevitably leads to the leak of information about the system under control, whose \textit{purity} typically decreases as it is driven to a mixed state.  

\begin{figure}[b]
\includegraphics[width= \columnwidth]{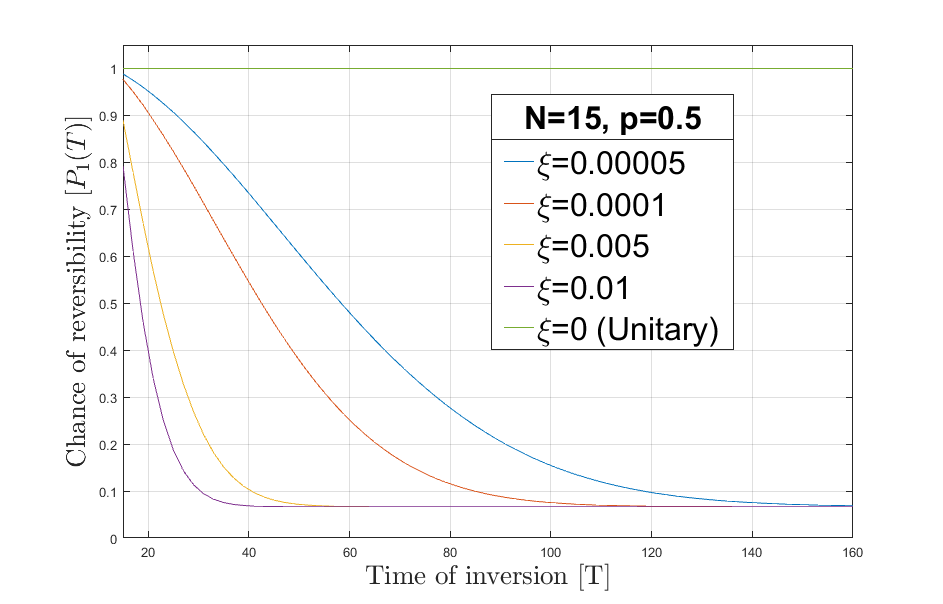}\\
\includegraphics[width= \columnwidth]{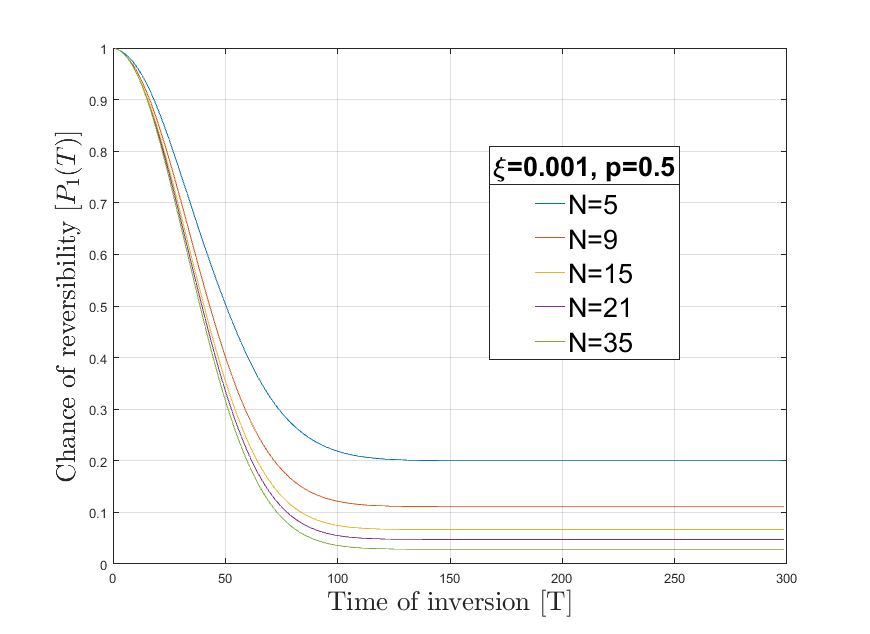}
\caption{Reversibility in the presence of pure dephasing. Upper panel: the probability $P_1$ of inverting the dynamics and end in the initial state $\rho_0$, as a function of the total evolution time $T$, for different values of the dephasing parameter $\xi$. Lower panel: same as above, for different lengths $N$ of the chain.}
\label{fig:dephasing}
\end{figure}

\begin{figure*}[t]
\includegraphics[width=\columnwidth]{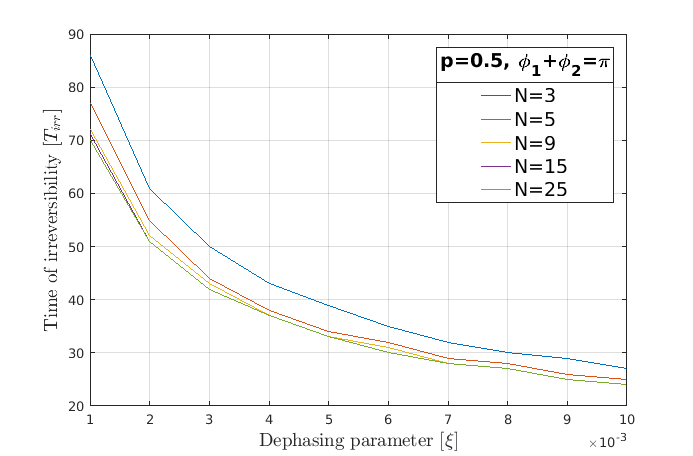}
\includegraphics[width=\columnwidth]{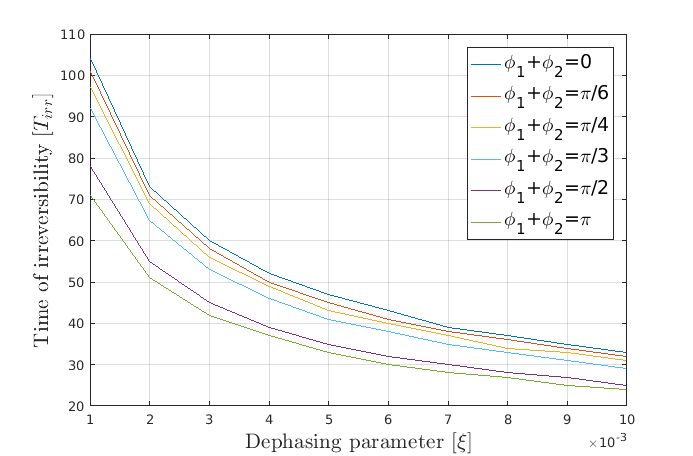}
\caption{Characteristic time for the occurrence of irreversibility of the system $T_{\rm irr}$ in the presence of pure dephasing. Left panel: $T_{\rm irr}$ as a function of the dephasing parameter $\xi$ for different legnths $N$; right panel: same as above, but for different phases $\phi_1+\phi_2$.}
\label{fig:time}
\end{figure*}

In the analysis and simulation of the properties of a many body quantum system the model of Quantum Cellular Automata (QCA) encases a privileged role \cite{bib:schumacher,watrous}. Its simple definition, in fact, subsumes a rich variety of possible, often complex, behaviours, and in recent years allowed the possibility to embrace within the same framework many different physical systems, from the simulation of quantum gases and quantum walks \cite{Aharonov} to the analysis of biological processes such as photosyntheses and eye vision \cite{bib:biology}. Furthermore these models have been proven to be capable of simulating a Quantum Turing Machine \cite{Raussendorf,Vollbrecht,Nagaj,Arrighi12,Dariano} while also being an ideal logical structure to construct causal quantum  theories \cite{Lloyd,Arrighi,Mosco}. Reversible cellular automata form a natural model of reversible computing. Besides, quantum cellular automata require reversibility in order to simulate properly Schr\"odinger equation, or any model in general that requires the conservation of physical quantities.

In the literature there has been a considerable effort in proofs and simulations that account for reversibility in both CA and QCA. So far, however, tests on a QCA coupled with a noisy environment are still missing. This is actually the most realistic and interesting case since any actual physical process will be affected by noise in some form, compromising its unitary evolution and hence its reversibility. For practical purposes, it is thus necessary to comprehend and control the role of quantum noise on the reversibility of the system. In order to investigate the properties of reversibility in quantum mechanical structures, we will make use of a specific model of noisy QCA developed by Avalle and Serafini \cite{Avalle1,Avalle2} for one-dimensional networks. Our studies illustrate how the system looses the possibility of inverting its evolution as the strength of the noise increases.
 

Despite the multitude of definitions used in literature to implement  QCA, the model has four main features dating back to the early proposal of John Von Neuman~\cite{bib:vonNeu1}: \textit{Discreteness:} the automaton consists in a quantum lattice, in which every site is an independent physical system with a finite set of states of a Hilbert space. The state of each site evolves in discrete time steps through the application of a specific rule, which is usually taken to be unitary, but can be generalized to accomodate the presence of noise. \textit{Homogeneity:} the rule is applied to every site at each time step and must be independent of lattice and time translations. \textit{Locality:} information must travel with finite speed through the lattice, hence the rule is applied only to a well specified neighborhood of each cell and its evolution at each step is independent from the rest of the lattice. \textit{Causality:} the evolution is deterministic, in the sense that knowing the configuration of the lattice at some time step it is possibile to predict completely the configuration in the next step.
In order for the update rule to be local and causal, it is essential to provide a convenient partitioning scheme, in which the lattice is divided in partitions of non-interacting neighbourhood blocks and the rule is applied to every block of a partition before considering the subsequent partition \cite{Schiff}. 

Quantum mechanics simplifies the problem of reversibility in cellular automata. In the classical case, in fact, the inverse of a CA is again a CA, provided that the rule is locally invertible, but it is a highly nontrivial matter to determine the neighborhood scheme of the inverse, which can be much larger than the neighborhood of the automaton itself. A general theorem on QCA, proved by Schumacher, ensures that \textit{any quantum cellular automaton is structurally reversible, the inverse of a nearest neighbor QCA exists, and is a nearest neighbor QCA}~\cite{bib:schumacher}.  The theorem is based on the Margolus neighborhood scheme, which gives a privileged role to partitioned QCA \cite{Margolus}. In order to invert time in a block partitioned QCA it is simply necessary to apply the inverse rule from the last block of the last partition to the first block of the first partition. 

\begin{figure*}[t]
\includegraphics[width= \textwidth]{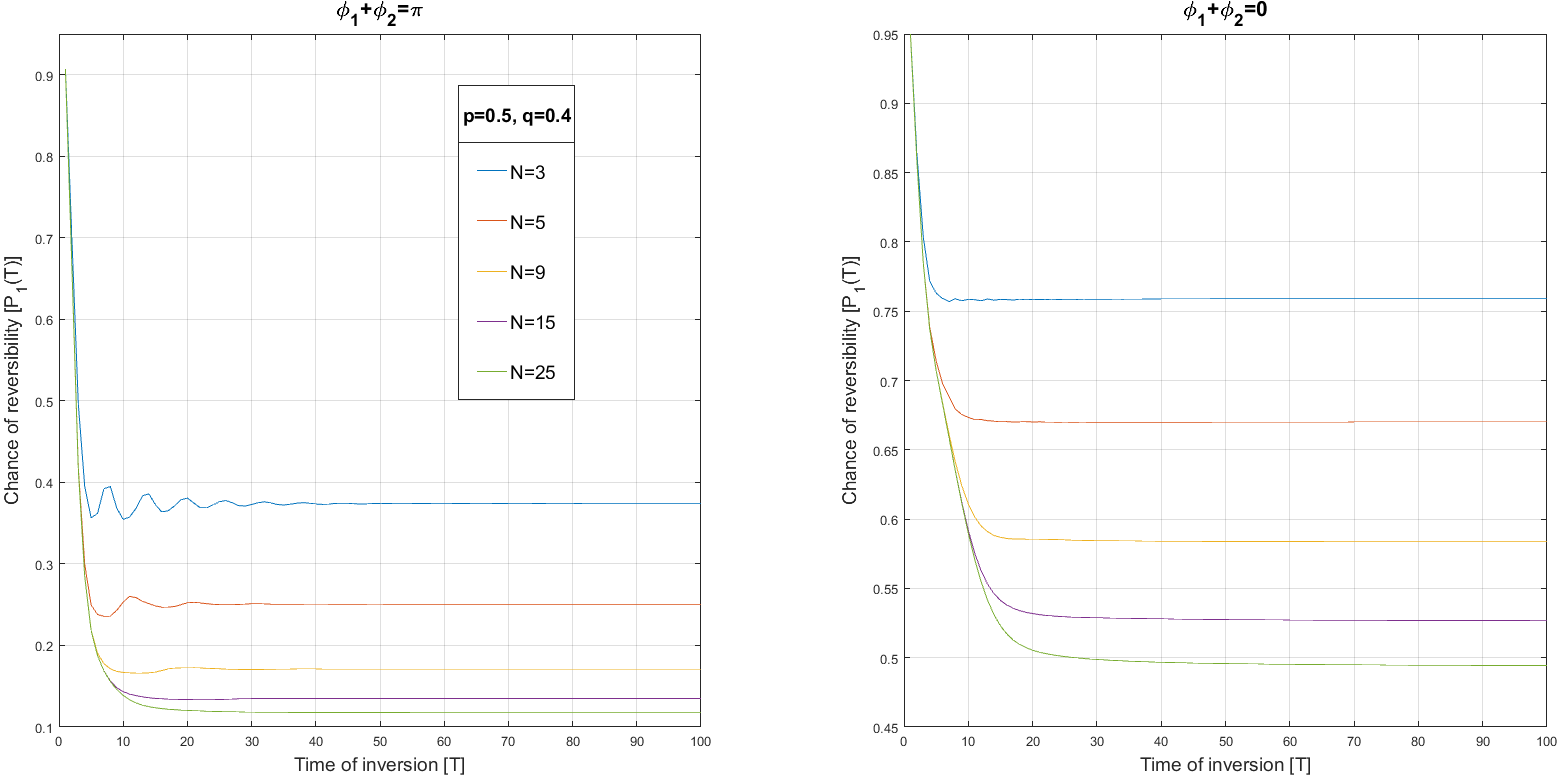}
\caption{Characteristic time for the occurrence of irreversibility of the system $T_{\rm irr}$ in the presence of pure dephasing. Left panel: $T_{\rm irr}$ as a function of the dephasing parameter $\xi$ for different legnths $N$; right panel: same as above, but for different phases $\phi_1+\phi_2$.}
\label{fig:damping}
\end{figure*}

Our QCA is constituted by a chain of $N$ sites, each being in either a single excited state or in a ground state, in a linear arrangement. The chain is initialised with an excitation on the first site, a pure state we indicate as $\rho_0$; the rule that transfers the excitation across sites consider a unitary coupling between the excited levels of the site $2n+1$ and that of the site $2n+2$, tensorised across all possible pairs so considered: these constitute two-site neighbourhood blocks. We will consider its parametrisation as: 
\begin{equation}
U =\frac{1}{\sqrt{1-p+q}}\left( 
\begin{matrix}
\sqrt{1-p} &  \sqrt{q} \,e^{i\phi_2} \\
\sqrt{q} \,e^{i\phi_1} &  -\sqrt{1-p} \,e^{i(\phi_1+\phi_2)}
\end{matrix}
\right),
\label{eqn:U}
\end{equation}
with $0\leq p,q \leq 1$ can be interpreted as the probabilities of moving the excitation to the site to the left of to the right site respectively. The partition is then shifted by one position in the chain, realising the coupling between the site $2n+2$ and $2n+3$. This concludes the elementary time step for the evolution of the QCA. Since excitations are not created nor destroyed at any time, the state of the QCA always belongs to the single-excitation sector. In addition to this unitary part, we also consider dissipative processes, dephasing and amplitude damping defined on the same two-level manifold as the unitary. These are combined with the unitary, leading to a generalised dynamics for the entire chain in the form:
\begin{equation}
\rho(t+1)=\Omega_{\xi,\eta}(\rho(t))=\Omega_{\xi,\eta}^{t+1}(\rho_0),
\end{equation}
where $\rho(t)$ is the density matrix of the whole system at the discrete time $t$, $\xi$ is the parameter quantifying dephasing, and $\eta$ is the parameter quantifying amplitude damping; we have that $0\leq\xi,|\eta|\leq1$, where 1 denotes a fully decohered process, and 0 a noiseless evolution~\cite{Avalle1,Avalle2,MikenIke}. In particular, the amplitude damping parameter is linked to those in the unitary \eqref{eqn:U} as $\eta=p-q$ \cite{note}.

The evolution operator $\Omega_{\xi,\eta}$ respects the partitioning in non-interacting neighbourhoods~\cite{Avalle1}, and contains both unitary and dissipative contributions. The latter cannot be inverted, thus we will introduce, as the `Schumacher inverse', the operation $\tilde\Omega_{\xi,\eta}$, defined as the one with the unitary $\Omega_{0,0}$ inverted following Schumacher's construction, 
followed by phase damping and amplitude damping with the same strengths $\xi$ and $\eta$ as in the original automaton. In order to quantify the degree of reversibility, we let the system evolve for $T/2$ time steps by $\Omega_{\xi,\eta}$, then invert the dynamics by applying $\tilde\Omega_{\xi,\eta}$ for the same number of steps; finally we consider the probability that the system ends in its initial state: $P_1(T) =\Tr{\left[\rho_0\, \tilde\Omega_{\xi,\eta}^{T/2}\left(\Omega_{\xi,\eta}^{T/2}(\rho_0)\right)\right]}$. 

Fig.~\ref{fig:dephasing} summarise the behaviour of the reversibility in the presence of dephasing. This rapidly drives the system towards irreversibility, leading it to the completely mixed state in the single-excitation sector. This can be easily verified by first noticing that the identity matrix ${\bf I}/N$ is a fixed point of the evolution since $\Omega_{\xi,0}({\bf I}/N)={\bf I}/N$. Further, this is the only fixed point, as implied by Banach's fixed point theorem: considering the trace distance ${\mathcal D}(\rho_1,\rho_2)$ between two arbitrary states in the single-excitation sector, it can be proved that the evolution map is contractive~\cite{MikenIke}:
\begin{equation}
{\mathcal D}\left( \Omega_{\xi,0}(\rho_1), \Omega_{\xi,0}(\rho_2)  \right)\leq \epsilon\, {\mathcal D}(\rho_1,\rho_2)
\end{equation}
where $0\leq\epsilon<1$. Consequently, there exist a single fixed point, represented by the completely mixed state. Once the system reaches this maximum entropic state, we witness its thermal death, thus the QCA is not anymore able to perform any computation or to transmit information: its arrow of time does not have a direction, since it is completely still. This condition is rapidly obtained, even in the presence of small levels of noise.

After a characteristic time, the probability of ending in the original state $\rho_0$ equals $1/N$, {\it i.e} the one obtained for a completely mixed state (Fig.~\ref{fig:dephasing}). We can introduce an operative measure for an inversion time $T_{\rm irr}$ as the one for which $P_1(T_{\rm irr}) \simeq 1/N$, to within a numerical threshold $\delta$ depending on the accuracy of the simulation; in our case, we choose $\delta=10^{-4}$, the corresponding results are shown in Fig.2. The time at which reversibility is completely lost depends loosely on the size of the chain: small-size effects ensure more robustness, however for long chains the system behaves very similarly.  Remarkably, the phases in the unitary $U$ in \eqref{eqn:U} set different limits to the reversibility, although in the pure limit $\xi=0$, perfect reversibility is clearly always achieved. It has been observed that the condition $\phi_1{+}\phi_2{=}0$ can lead to localisation of the excitation, due to interference effects~\cite{Avalle1}; such localisation is not observed when $\phi_1{+}\phi_2{=}\pi$. This leads to a faster loss of reversibility, when compared to the occurrence of localisations; these, in fact, are partly protected against the noise, and retain the capability of walking back on its steps for longer times.

Decoherence under amplitude damping presents quantitative and qualitative difference with respect to the previous case of dephasing. However, as illustrated in  Fig.~\ref{fig:damping}, the presence of localisation is always relevant to the reversibility. In their absence, the loss reversibility occurs with a largely size-independent characteristic time. However, the final state of the evolution does not correspond to the completely mixed state, due to the population transfer operated by the noise: the probability $P_1(T)$ is always larger than the one for ${\bf I}/N$. For long times, there is an equilibrium between noisy and unitary transfer mechanisms. Small-size effects are manifested as small oscillations, which are rapidly damped if the size is increased. Localisation changes this behaviour: the probability $P_1(T)$ is now considerably above the mixed-state limit, even for long chains. This is due to the fact that is is easier for the noise to pump population into the first site starting from a localised state, than it is from a delocalised situation. Qualitatively, the increase in the long-time value of $P_1(T)$ can be understood by noticing that amplitude damping in part freezes the evolution of the system by forcing it to remain in the initial state: the system retains some of its reversibility since it has not actually undergone any evolution.

In conclusion, we have introduced a notion of reversibility for noisy quantum cellular automata, and used it to inspect how noise processes affect the probability of restoring the initial state following inversion of the evolution. We have used it to define a characteristic time for the system to achieve a steady state, and we have illustrated how different classes of noise process introduce qualitative difference in the reversibility of such steady states. Furthermore, we drew considerations on the effects of excitation localisations in the reversibility that become manifest only in the presence of noise.

{\it Acknowledgements.} We thank P. Mataloni, M. Paris and M. Avalle for discussion and encouragement.


\begin{thebibliography}{99}
\bibitem{note} For $\eta<0$ one can still define an amplitude damping channel as the 'swapped' version of the conventional definition.
\end{thebibliography}
\end{document}